\documentclass{article}
\usepackage{graphicx,latexsym}

\def\b{\beta}
\def\dl{\delta}
\def\s{\sigma}

\def\lam{\lambda}
\def\Lam{\Lambda}

\def\bg{\bar{g}}

\def\bDelta{\bar{\Delta}}
\def\P{{\rm P}}
\def\QG{{\rm QG}}
\def\D{{\rm D}}

\def\sq{\sqrt}
\def\e{\hbox{\large \it e}}

\def\bx{{\bf x}}

\def\lap3{~| \!\!\! \partial^2}
\def\dlap3{~| \!\!\! \partial^4}
\def\dH{{\dot H}}
\def\ddH{{\ddot H}}
\def\dddH{\stackrel{...}{H}}

\def\bb{\begin{equation}}
\def\ee{\end{equation}}
\def\bba{\begin{eqnarray}}
\def\eea{\end{eqnarray}}

\begin{document}

\begin{titlepage}

\begin{flushright}
{\sc May 2007}
\end{flushright}

\begin{center}
{\Large {\bf Analyzing WMAP Observation \\ by Quantum Gravity}}
\end{center}

\vspace{5mm}

\begin{center}
{\sc Ken-ji Hamada$^1$, Shinichi Horata$^2$, Naoshi Sugiyama$^3$ \\ and Tetsuyuki Yukawa$^4$}
\end{center}

\begin{center}
{}$^1${\it Institute of Particle and Nuclear Studies, KEK, Tsukuba 305-0801, Japan} \\
{}$^1${\it Department of Particle and Nuclear Physics, The Graduate University for Advanced Studies (Sokendai), Tsukuba 305-0801, Japan} \\
{}$^{2,4}${\it Hayama Center for Advanced Research, The Graduate University for Advanced Studies (Sokendai), Hayama 240-0193, Japan} \\
{}$^3${\it Graduate School of Science, Nagoya University, Nagoya 467-8602, Japan}
\end{center}

\begin{abstract}
The angular power spectra of cosmic microwave background are analyzed under the light of the evolutional scenario of the universe based on the renormalizable quantum theory of gravity in four dimensions. The equation of evolution is solved numerically fixing the power law spectrum predicted by the conformal gravity for the initial condition. The equation requires to introduce a dynamical energy scale about $10^{17}$GeV, where the inflationary space-time evolution makes a transition to the big-bang of the conventional Friedmann universe. The quality of fit to the three-year data of WMAP implies the possibility to understand the observation by quantum gravity.

\vspace{5mm} 

\end{abstract}
\end{titlepage}

Observations of the cosmic microwave background (CMB) anisotropies by the cosmic background explorer (COBE) \cite{cobe} and the Wilkinson microwave anisotropy probe (WMAP) \cite{wmap,wmap3,wmap3p} have established the inflationary scenario of the universe. Various cosmological parameters have been determined precisely based on the cosmological perturbation theory \cite{bardeen,ks,hs,ll}. In those analyses, however, the primordial spectrum close to that of the Harrison-Zel'dovich \cite{hz} is assumed as the initial condition of the big bang scenario of the universe. One of the important problems we wish to discuss in this letter is the dynamics to produce such a primordial spectrum in terms of the fundamental point of view rather than introducing an ambiguous field from phenomenological purpose.

It has been recognized that before the big bang there is an inflationary expanding epoch which solves the horizon and the flatness problems \cite{guth,starobinsky}. The inflation naively says that the universe grew about the order of $10^{60}$ from the birth to the present, which implies that the size of the Hubble distance today was placed within the Planck length at the Planck time. Thus, it is natural to consider that the primordial spectrum was originated from quantum fluctuations of space-time and these Planck scale phenomena are recorded as the CMB anisotropies observed by COBE and WMAP \cite{hy}.

Since the observed spectrum of constant $l(l+1)C_l$ suggests an almost scale invariant primordial spectrum for the large angle correlation, it is natural to expect that initial fluctuations are generated by a dimensionless field, and the only candidate is the metric field within known physical fields in the microscopic scale.

The inflationary scenario induced by quantum effects of gravity was first proposed by Starobinsky in 1979 \cite{starobinsky}. His idea has an advantage that it is not necessary to introduce any additional fields. At that time, however, the idea cannot explain how the inflation terminates to the big bang after expansion with sufficient number of e-foldings and why tensor fluctuations are less dominated than scalar ones in the very early universe while quantum gravity effects are so strong.

The renomalizable quantum conformal gravity we employ here can explain these problems in terms of the asymptotically free property of the traceless tensor mode in the metric field whose dynamics is governed by the Weyl action. It suggests that there is a dynamical energy scale separating between quantum and classical space-time, and beyond this energy scale, an inflationary quantum space-time is realized, and then the conformal-mode fluctuation dominates. In this way, Starobinsky's idea of inflation is revived on the foundation of the modern field theory.

Quantum fluctuations of the conformal mode are getting small during the inflation. Applying the linear perturbation theory about the inflationary background, we compute the transfer functions from the Planck time to the big bang which took place at the dynamical scale \cite{hhy}. We then calculate the multipole components of the CMB angular power spectrum comparing with the WMAP data.

\paragraph{Evolutional Scenario of The Universe}
Renormalizable quantum gravity \cite{hamada02,nova} is defined by the four-derivative conformal invariant actions, $(-1/t^2)\sq{-g}C_{\mu\nu\lam\s}^2$ and $- b \sq{-g} G_4$, where $C_{\mu\nu\lam\s}$ is the Weyl tensor and $G_4$ is the Euler density, in addition to the Einstein-Hilbert action and conformal invariant matter fields. The dimensionless coupling constant $t$ in the Weyl action is introduced to take care of the traceless tensor mode at the short distance scale, while the constant $b$, which is introduced to renormalize divergences proportional to the Euler density, is not an independent coupling constant because the Euler term does not have the dynamical components. Quantization is carried out perturbatively about conformal flat space-time with the vanishing Weyl tensor, and thus the metric field is expanded as $g_{\mu\nu}=\e^{2\phi}\bg_{\mu\nu}$ with $\bg_{\mu\nu}=\eta_{\mu\nu}+h_{\mu\nu}+\cdots$, where $\phi$ and $h_{\mu\nu}$ are the conformal mode and the traceless tensor mode, respectively.

The renomalized coupling, $t_r$, for the traceless tensor mode is shown to be asymptotic free, whose beta function was computed in \cite{ft,hamada02} as $\b=-\b_0 t_r^3$ with $\b_0 >0$. This justifies the perturbative treatment for this mode, and also implies the existence of a dynamical scale $\Lam_\QG$, where the running coupling constant is written as
\begin{equation}
   1/t_r^2(p)=\b_0 \ln(p^2/\Lam^2_\QG)
\end{equation}
for a physical momentum $p$. The asymptotic freedom yields that the coupling constant is getting small at very high energies, and configurations with the vanishing Weyl tensor are chosen quantum mechanically that prohibit a singular configuration with a divergent Riemann curvature tensor.

The conformal mode is quantized non-perturbatively so that the conformal invariance becomes exact when the traceless-mode coupling constant vanishes at very high energies. The dynamics of the conformal mode is induced from the measure as the Wess-Zumino action of conformal anomaly, known as the Riegert action, containing the kinetic term $(-b_1/8\pi^2)\times \sq{-\bg}\phi\bDelta_4\phi$ \cite{riegert,am,amm92,hamada99}, where $\bDelta_4$ denotes the fourth-order conformal invariant operator defined on $\bg_{\mu\nu}$. The coefficient has been computed within the lowest order as $b_1=(2N_{\rm X}+11N_{\rm W}+124N_{\rm A})/720+769/180$ \cite{amm92}, where $N_{\rm X}$, $N_{\rm W}$ and $N_{\rm A}$ are the numbers of conformal scalar fields, Weyl fermions and gauge fields, respectively.

Evolution of the early universe is divided into three stages by two mass scales, the reduced Planck scale $M_\P=1/\sq{8\pi G}$ and the dynamical scale $\Lam_\QG$, ordered as $M_\P \gg \Lam_\QG$ \cite{hy}. At very high energies beyond the Planck scale, the space-time is dominated by the quantum fluctuation with exact conformal invariance. The symmetry begins to be broken about the Planck scale, toward the stage of inflationary expanding universe with the Hubble constant $H_\D=\sq{8\pi^2/b_1}M_\P$. The running coupling gradually increases during inflation, and it diverges at the dynamical scale $\Lam_\QG$. It is the period that the conformal invariance is completely broken and correlation length becomes short to result the emergence of classical space-time.

The dynamics of the inflationary phase before the big bang is effectively described by including corrections of the traceless-mode coupling constant to the Wess-Zumino action such as $b_1(1-a_1 t_r^2+\cdots)=b_1B_0(t_r)$ with $a_1 > 0$. The higher order effects are taking into account by a resummation form $B_0=1/(1+a_1 t_r^2/\kappa)^\kappa$, where $\kappa$ is a parameter that lies in the range $0 < \kappa \leq 1$. Under the spirit of the mean-field approximation, we simplify the momentum dependence of the running coupling by its time-dependent average: replacing the physical momentum by the inverse of the proper time $\tau$ as $1/t^2_r(\tau)=\b_0\ln(1/\tau^2\Lam^2_\QG)$. It shows that the running coupling diverges at the dynamical time scale $1/\Lam_\QG ~(=\tau_\Lam)$, and then the dynamical factor $B_0$ vanishes indicating the transition from the conformal gravity to the Einstein gravity. In this way, we obtain the evolutional homogeneous equations of motion \cite{hhy,nova},
\begin{equation}
     B_0(\tau) \left( \dddH +7H\ddH +4\dH^2 +18H^2\dH +6H^4 \right)
     -3 H_\D^2 \left( \dH +2H^2 \right) =0 
     \label{evolution}
\end{equation}
and the conservation equation 
\begin{equation}
     B_0(\tau) \left( 2H\ddH -\dH^2 +6H^2\dH +3 H^4 \right) -3 H_\D^2 H^2
        + 8\pi^2\rho/b_1 =0,
\end{equation}
where $H$ is the Hubble parameter defined by $H={\dot a}/a$ and $a=\e^{\phi}$ is the scale factor. The dot denotes the derivative with respect to the proper time and $\rho$ is the matter density.

The space-time initially evolves in an inflationary expansion with $H=H_\D$. The number of e-foldings from the Planck time $1/H_\D ~(=\tau_\P)$ to $\tau_\Lam$ is approximately given by the ratio of two mass scales: ${\cal N}_e =\log[a(\tau_\Lam)/a(\tau_\P)] \sim H_\D/\Lam_\QG$, which will be set about $H_\D/\Lam_\QG=60$ as popularly accepted. The coefficient of the Wess-Zumino action $b_1$ is taken as $15$ and $20$ for the analyses to compare to the WMAP3 data in the following. The dynamical scale is then given as $\Lam_\QG \simeq 10^{17}$GeV.

\begin{figure}
\begin{center}
 \begin{tabular}{cc}
 \resizebox{60mm}{!}{\includegraphics{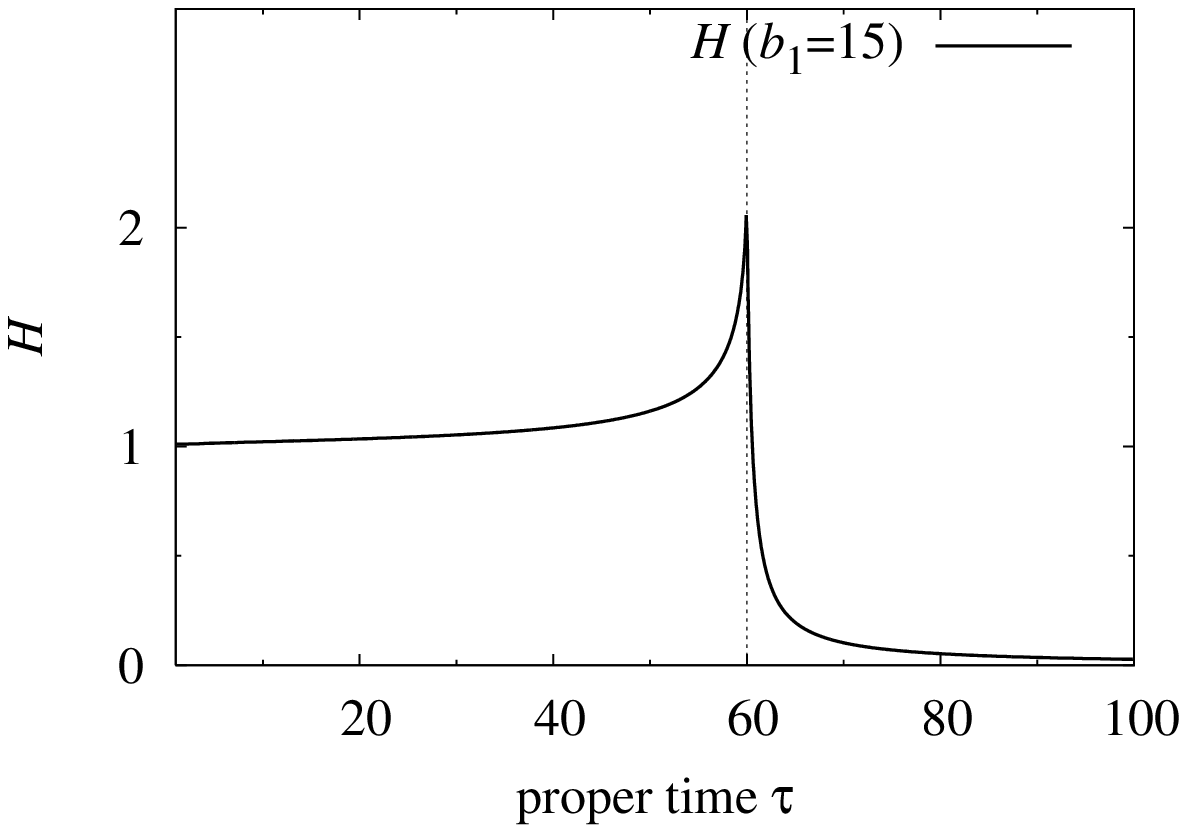}} &
 \resizebox{60mm}{!}{\includegraphics{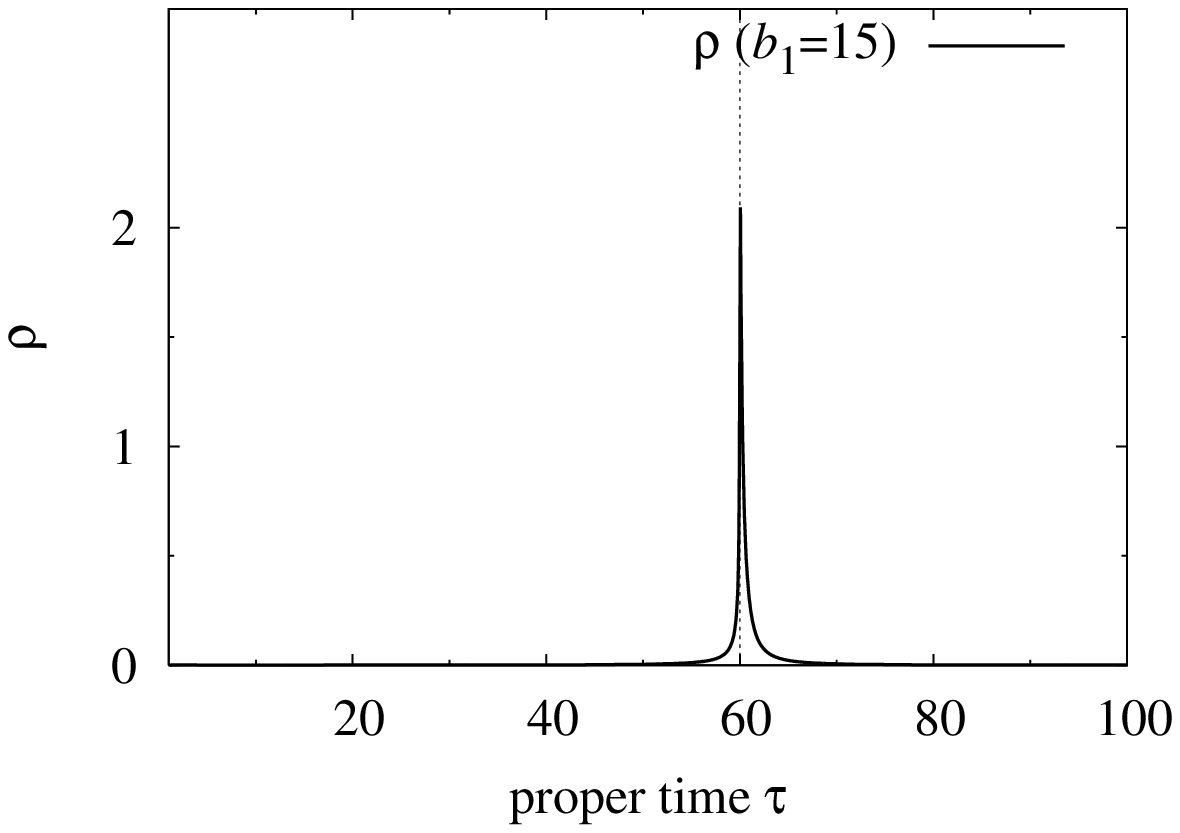}} \\
 \end{tabular}
\end{center}
\caption
{\label{fig 1} The time evolution of $H$ and $\rho$ for $b_1=15$ and $H_\D/\Lam_\QG=60$, where $H_\D$ is normalized to be unity.}
\end{figure}

The other parameters in the model are rather insignificant. Since they depend on the non-perturbative dynamics of the traceless mode, they are chosen phenomenologically as $\beta_0/b_1=0.06$, $a_1/b_1=0.01$, and $\kappa=0.5$. With these combinations of parameters the homogeneous equation (\ref{evolution}) and the evolution equations for scalar and tensor fluctuations, which will appear later, preserve their forms independent of $b_1$. The number of e-foldings is then computed to be ${\cal N}_e=65.0$. The solution of the evolution equations is depicted in figure \ref{fig 1} for the case of $b_1=15$, where $H_\D$ is normalized to be unity. The sharp increase of the matter density at the transition point indicates the big bang where energies stored in the conformal mode shift to the matter degrees of freedom.

Below the energy scale $\Lam_\QG$, the Einstein action becomes dominant and the space-time makes transition to the classical phase. This phase is described by the low energy effective theory of gravity expanding in derivatives of the metric field \cite{hhy} in an analogy to the chiral perturbation theory for QCD. For simplicity, we connect the conformal universe to the Friedmann universe at the transition point. This simplification will not cause much effects in later discussions because we consider fluctuations with the size of the Planck length at the Planck time. Therefore, at the transition point the size is extended much more than the correlation length, and the patterns of spectra are insignificant on the dynamics and the parameters $\b_0$, $a_1$, $\kappa$ employed at the transition.

\paragraph{Primordial Spectra}
Since the inflationary solution is stable, gravitational fluctuations about this solution get smaller during the inflation. Let us first give a rough estimation for the amplitude of scalar fluctuation considering a dimensionless contrast of the scalar curvature fluctuation, $\dl R/R$. The denominator is the curvature of inflationary background, which is $12H_\D^2$ for de Sitter curvature with $H=H_\D$. Since the curvature has two derivatives, the curvature fluctuation would be order of the square of the energy scale, $E^2$. Hence, the amplitude of scalar fluctuation is estimated to be $\dl R/R \sim E^2/12H_\D^2$. This implies that we can apply the linear perturbation theory for the density fluctuation under the inflationary background in the period from the Planck time $\tau_\P$ to the dynamical time $\tau_\Lam$. At the dynamical energy scale, the amplitude of the scalar fluctuation is estimated as $\Lam_\QG^2/12H_\D^2 \sim 1/12{\cal N}_e^2$, which gives the magnitude about the order of $10^{-5}$ similar to the observations.

For the scalar fluctuation we compute the evolutions of the so-called Bardeen potentials defined by $ds^2=a^2[-(1+2\Psi)d\eta^2 +(1+2\Phi)d\bx^2]$. The coupled equations for the evolutions of these fields in the inflationary background have been derived in \cite{hhy}. The initial configuration at the Planck time is given by $\Phi=\Psi$ because of the dominance of conformal-mode fluctuations at very high energies, while at the transition point the dynamics requires the configuration satisfying $\Phi=-\Psi$. The scalar spectrum defined by the two-point correlation of the Bardeen potential is given by the quantum conformal gravity as 
\begin{equation}    
       P_s^{\rm pl}(k)=A_s(k/m)^{(n_s-1)}
       \label{scalar}
\end{equation}
at the Planck time, where $k$ is a spatial comoving momentum and $m=a(\tau_\P)H_\D$ is the comoving Planck scale at the Planck time. Since this scale appears in the evolutional equations of fluctuations, it is a dynamical parameter to determine the pattern of the spectra at the transition point. The dimensionless amplitude $\sq{A_s}$ is given by the order of $10^{-1}$ obtained by substituting $E \simeq H_\D$ into the expression of curvature fluctuation estimated above. The scalar spectral index is given by the anomalous dimensions of the scalar curvature \cite{amm-cmb,hy}, 
\begin{equation}
    n_s=5-8(1-\sq{1-2/b_1})/(1-\sq{1-4/b_1}).
\end{equation}
In the large $b_1$ limit, it approaches to the Harrison-Zel'dovich spectrum with $n_s=1+2/b_1+4/b_1^2+o(1/b_1^3)$.

The tensor spectrum at the Planck time is given by 
\begin{equation}
     P_t^{\rm pl}(k)=A_t (k/m)^{n_t}.
     \label{tensor}
\end{equation}    
Because of the asymptotic freedom for the traceless tensor mode, the amplitude is considered to be much smaller than the scalar amplitude $(A_t \ll A_s)$ and the index is given by $n_t=0$.

The transfer functions in the inflationary period are defined by $\Phi(\tau_\Lam,k)={\cal T}_s(\tau_\Lam,\tau_\P) \Phi(\tau_\P,k)$ for the scalar mode and $h^{\rm TT}_{ij}(\tau_\Lam,k)={\cal T}_t(\tau_\Lam,\tau_\P) h^{\rm TT}_{ij}(\tau_\P,k)$ for the tensor mode. The primordial spectra at big bang are obtained as $P_s(k)={\cal T}_s^2(\tau_\Lam,\tau_\P) P_s^{\rm pl}(k)$ and $P_t(k)={\cal T}_t^2(\tau_\Lam,\tau_\P) P_t^{\rm pl}(k)$.

\begin{figure}[t]
\begin{center}
\includegraphics[scale=0.9]{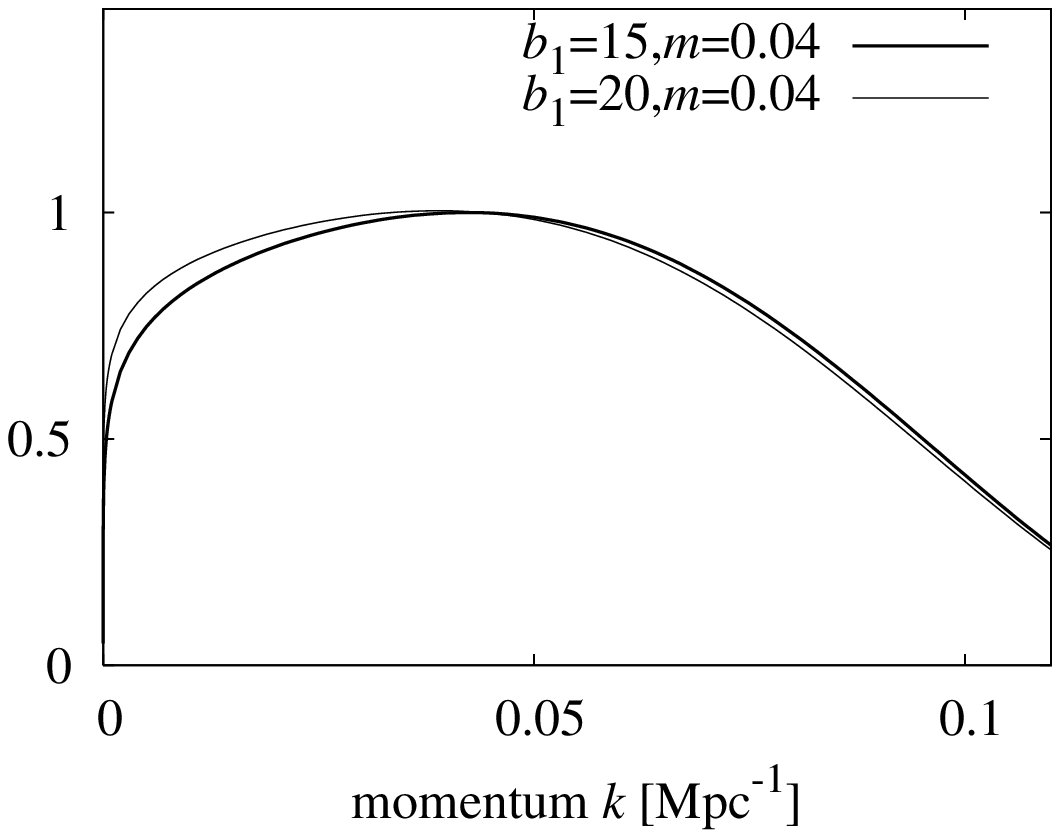}
\includegraphics[scale=0.9]{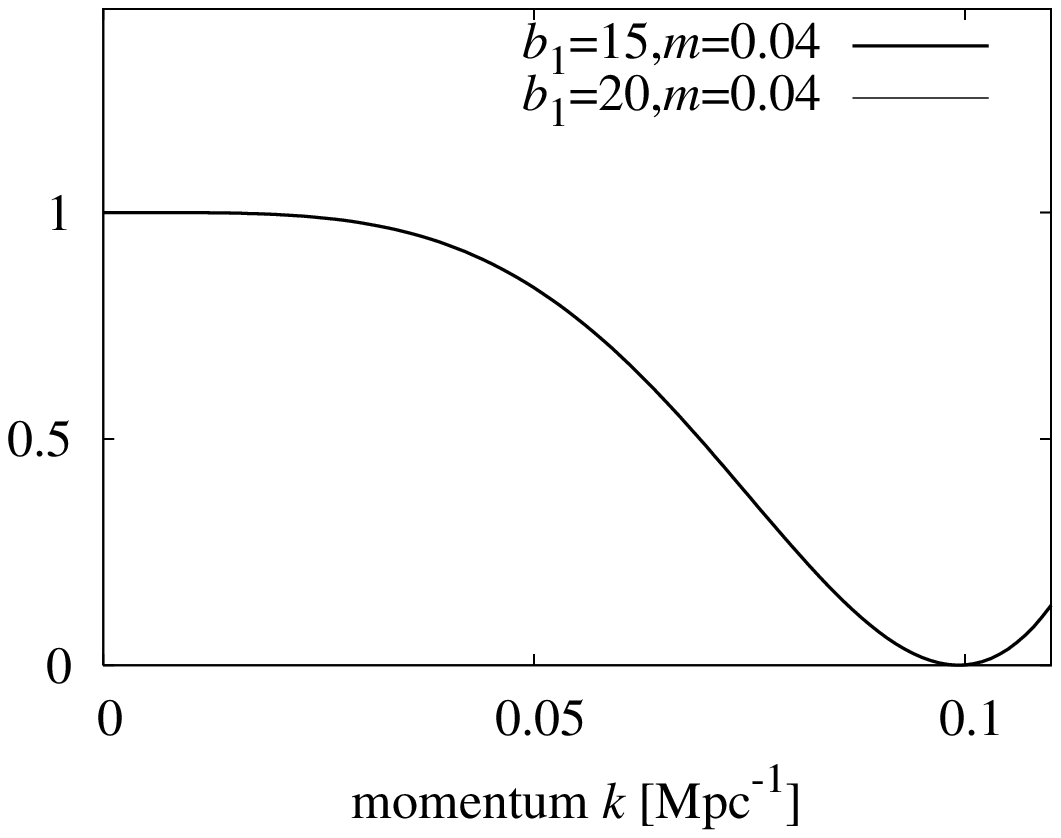} 
\end{center}
\caption{\label{fig 2} Patterns of the scalar and tensor spectra, $P_s$ and $P_t$, for $b_1=15$ (thick) and $20$ (thin). In each case, the simulation is carried out by $m=0.04$Mpc$^{-1}$, and the amplitude is normalized appropriately. For the tensor spectrum, there is no significant dependence on the value of $b_1$.
}
\end{figure}

The transfer functions are computed numerically, and it is shown that the amplitude of the scalar fluctuation gets smaller as estimated above, while the tensor fluctuation stays to be small \cite{hhy}. The scalar spectrum that is initially blue $(n_s>1)$ shifts to red in higher momentum regions with $k>m$. From this behavior we can fix the comoving Planck scale. Since the WMAP data favors the scalar spectral index less than one at $k=0.05$Mpc$^{-1}$, the value of $m$ is taken to be as small as $0.04$Mpc$^{-1}$. In figure \ref{fig 2} we show the result of simulation for the case of $b_1=15$ and $20$ with $m=0.04$Mpc$^{-1}$. This value of $m$ implies that the scale factor at the Planck time is $a(\tau_\P) \simeq 10^{-59}$ when we consider the size of Planck length at the Planck time grows up to the size $1/m$ today.

\paragraph{CMB Multipoles}

\begin{figure}
\begin{center}
\includegraphics[scale=1]{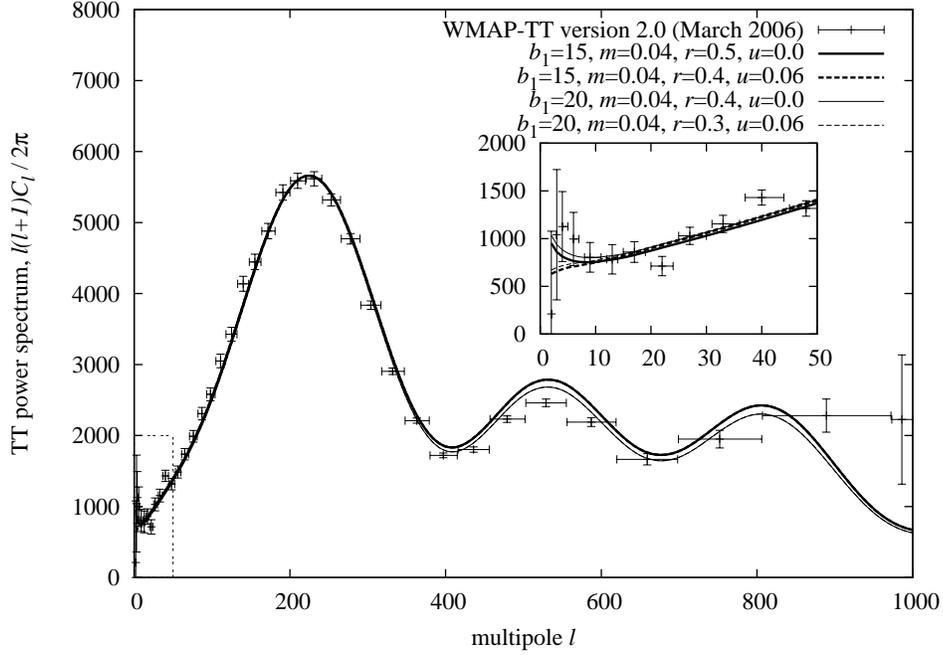}
\end{center}
\caption{\label{fig 3} The TT spectra for $b_1=15$(thick) and $20$(thin) with $m=0.04$Mpc$^{-1}$. The tensor-to-scalar ratio $r$ and the damping factor $u$ are determined to fit with the WMAP3 data. 
The amplitude is normalized at the first peak. The part at low-multipole components is enlarged and depicted to see the $r$ and $u$ dependence for each $b_1$, while there is no significant dependence on $u$ at high-multipole components. The other cosmological parameters are fixed to be the best fit values in \cite{wmap3} such as $\Omega_{\rm b}=0.041$, $\Omega_{\rm cdm}=0.205$, $\Omega_{\rm vac}=0.754$, $H_0=73.1$, $\tau_e=0.108$, $T_{\rm cmb}=2.726$, and $Y_{\rm He}=0.24$.
}
\end{figure}

\begin{figure}
\begin{center}
\includegraphics[scale=1]{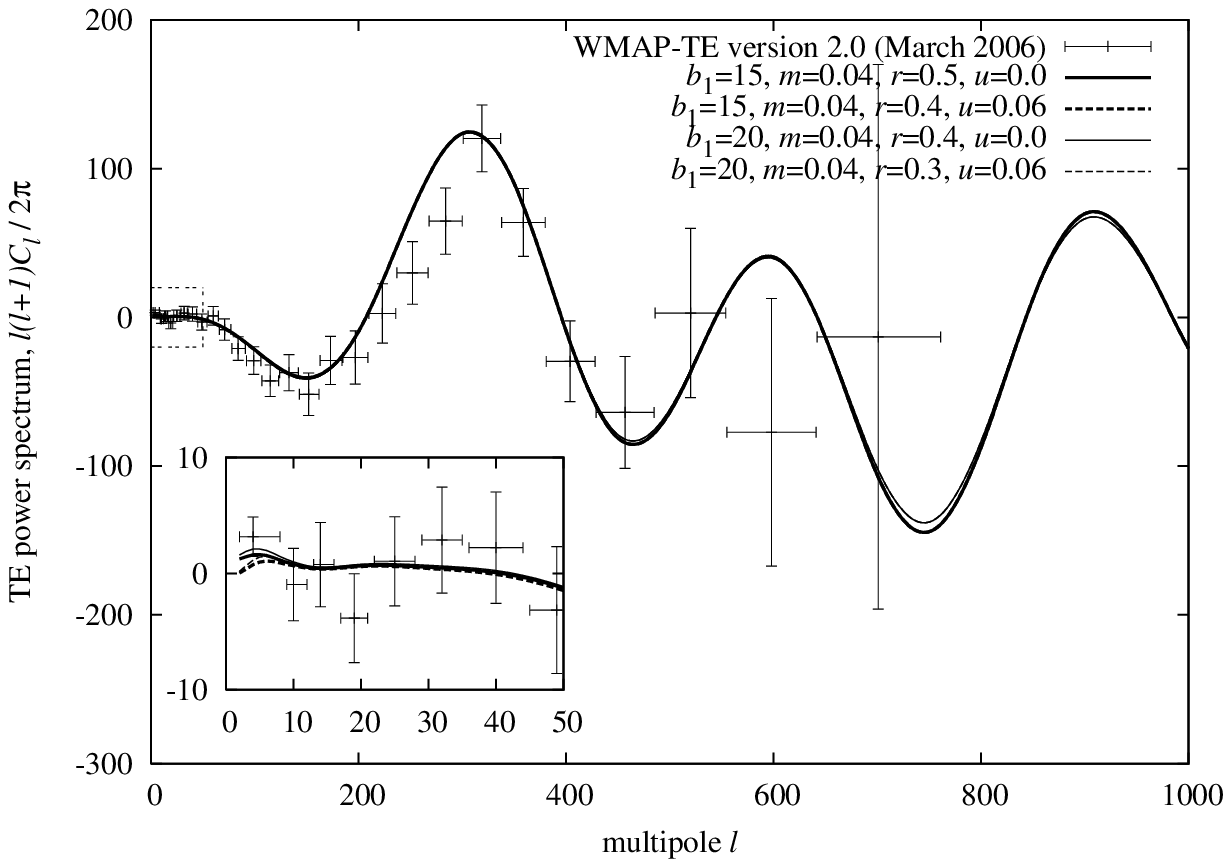}
\includegraphics[scale=1]{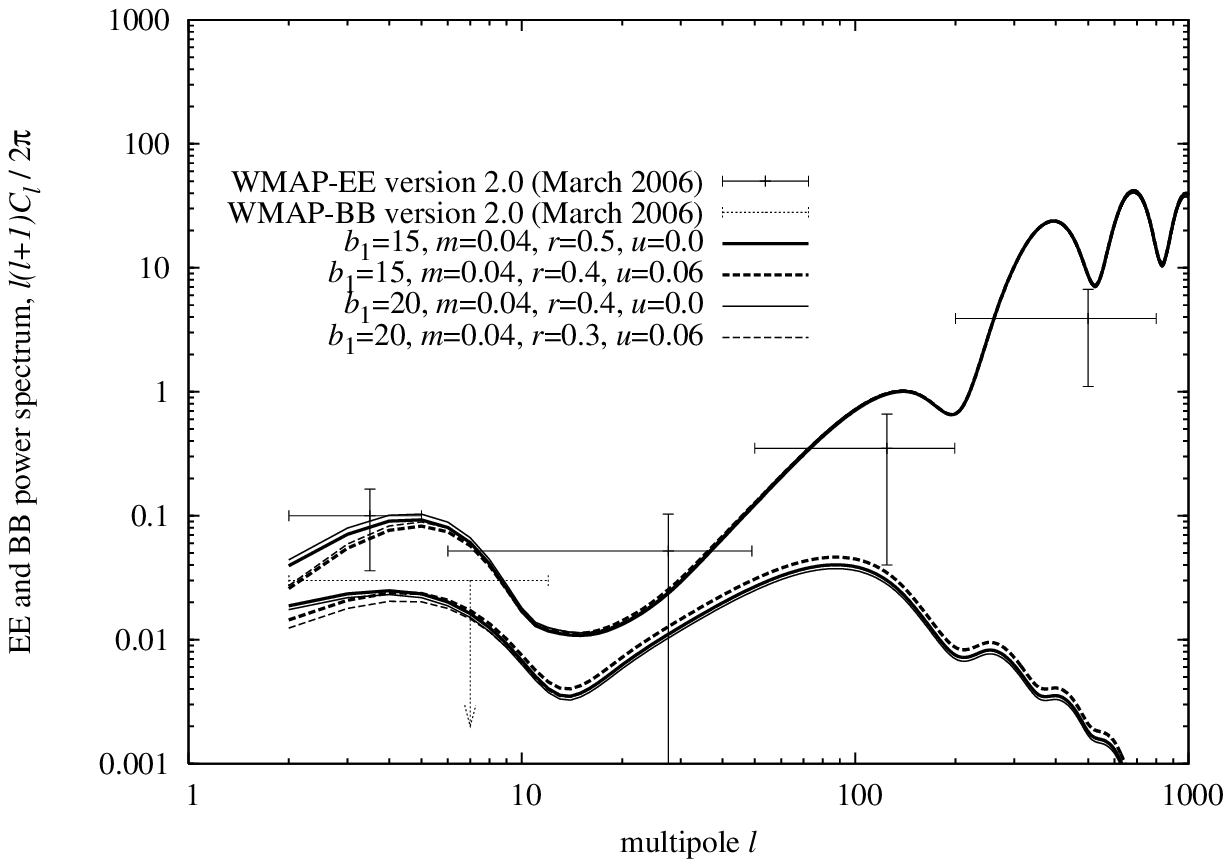}
\end{center}
\caption{\label{fig 4} The TE, EE, and BB polarization spectra.}
\end{figure}

The CMB angular power spectra are calculated using the cmbfast code \cite{cmbfast}. The initial conditions to be supplied are chosen from the transfer functions obtained at the transition time specified above. The coefficient of the Wess-Zumino action $b_1$ which depends on the matter contents is taken to be larger than that for typical GUT models\footnote{
For $SU(5)$, $SO(10)$, and $E_6$ models, $b_1 \simeq 9$, $12$, and $18$, respectively
} 
so that the scalar spectral index is close to one for the momentum region less than $m$. In figures \ref{fig 3} and \ref{fig 4} we show the numerical results of TT, TE, EE, and BB spectra for $b_1=15(n_s=1.15)$ and $20(n_s=1.11)$ together with the WMAP3 data. The other primordial parameters such as the tensor-to-scalar ratio $r$ and the amplitudes are adjusted to fit the observation data. In order to see the dependence of the primordial parameters, we employ the cosmological parameters to be the same as the best fit values of the analysis of WMAP \cite{wmap3}.

The suppression of the low multipole components may be explained as the consequence of the dynamical scale. If we wish to discuss it rigorously we need to evaluate the two-point correlation function for long-distance separation, which will involve non-perturbative dynamics of the traceless mode. Instead, we here simply give a phenomenological ansatz based on the dimensional analysis. The damping factor for a spatial separation far beyond the correlation length is assumed to have the following form: $k^2/(k^2+u\lam^2)$, where $\lam=a(\tau_\P)\Lam_\QG=m/60$ is the comoving dynamical scale at the Planck time. This factor should be multiplied to the initial spectra (\ref{scalar}) and (\ref{tensor}). 
The simulation results with the damping factor are depicted by the dashed lines in figure \ref{fig 3} and \ref{fig 4}. There is small improvement on the $l=2$ component in the TT spectrum, but it seems to suggest that we need further investigations on the dynamics of the two-point correlation requiring more sharp fall off.

For higher momentum region more than $k > 2m$, we need to study with considering non-linear effects of the evolution equation, because the fluctuation with high momentum beyond this range corresponds to that with the momentum far beyond the Planck scale initially. For the scalar fluctuation, the non-linearity of conformal field theory becomes significant until the amplitude reduces small enough to apply the linear approximation safely. We regard the non-linear effect hopefully sustains the amplitude so as not to fall off at high momentum region which is required by other observational results. As for the tensor fluctuations, the linear approximation is applicable in the high momentum region due to the asymptotically free property of this mode.

\paragraph{Conclusion}
The renormalizable quantum gravity based on the conformal gravity in four dimensions suggests that there appears a dynamical energy scale $\Lam_\QG$ about the order of $10^{17}$GeV, between the Planck scale and the GUT scale. This energy scale separates the conventional classical space-time from quantum space-time with conformal invariance. The big bang can be understood as a space-time transition where energies stored in extra degrees of freedom in higher-derivative gravitational fields shift to matter degrees of freedom.

The CMB angular power spectra are calculated on the basis of the quantum gravity. We numerically simulate evolutions of the scalar and tensor fluctuations during the inflationary period from the Planck time to the transition time, and by setting the computed spectra at the transition as the initial conditions of the cmbfast code we obtain the TT, TE, EE, and BB angular power spectra. The results fit to the WMAP3 data, which suggests that the renormalizable quantum gravity is one of the possibility to explain the origin of CMB anisotropies.

The condition that the dynamical energy scale is being lower than the Planck mass implies that quantum effects turn on much larger scale than the Planck length, and thus very high energy particle is dressed by quantum gravity and space-time is deformed locally to prevent the particle itself from forming a black hole. Such a quantum gravity effect about the dynamical scale might be observed in future experiments such as observations of gamma-ray bursts \cite{aemns}, in addition to CMB.


\newpage

\end{document}